\documentclass[prb,a4paper,showpacs,twocolumn,superscriptaddress,longbibliography]{revtex4-2}
\usepackage[utf8]{inputenc}

\usepackage{graphicx}
\usepackage{fullpage}
\usepackage{amsmath}
\usepackage{amssymb}
\usepackage{bm}
\usepackage{hyperref}
\usepackage{cleveref}
\usepackage{svg}
\usepackage{braket}

\usepackage{inputenc}

\DeclareMathOperator{\Tr}{Tr}

\begin{document}
  
\title{Drastic effect of weak interaction 
near special points in semiclassical multiterminal superconducting nanostructures}

\author{Janis Erdmanis} 
\affiliation{Kavli Institute of Nanoscience, Delft University of Technology, 2628 CJ Delft, The Netherlands}
\author{Árpád Lukács} 
\affiliation{Wigner RCP RMI, H1525 Budapest, POB 49}
\affiliation{Department of Theoretical Physics, University of the Basque Country UPV/EHU, POB 644, E-48080 Bilbao, Spain}
\author{Yuli Nazarov}
\affiliation{Kavli Institute of Nanoscience, Delft University of Technology, 2628 CJ Delft, The Netherlands}
\begin{abstract}
A generic semiclassical superconducting nanostructure connected to multiple superconducting terminals hosts a quasi-continuous spectrum of Andreev states.  The spectrum is sensitive to the superconducting phases of the terminals. It can be either gapped or gapless depending on the point in the multi-dimensional parametric space of these phases. Special points in this space correspond to setting some terminals to the phase 0 and the rest to the phase of $\pi$. For a generic nanostructure, three distint spectra come together in the vicinity of a special point: two gapped phases of different topology and a gapless phase separating the two by virtue of topological protection.

In this paper, we show that a weak interaction manifesting as quantum fluctuations of superconducting phases drastically changes the spectrum in a narrow vicinity of a special point. We develop an interaction model and derive a universal generic quantum action that describes this situation. The action is complicatesd incorporating a non-local in time matrix order parameter, and its full analysis is beyond the scope of the present paper. Here, we identify and address two limits: the semiclassical one and the quantum one, concentrating on the first-order interaction correction in the last case.

In both cases, we find that the interaction squeezes the domain of the gapless phase in the narrow vicinity of the point so the gapped phases tend to contact each other immediately defying the topological protection. We identify the domains of strong coupling where the perturbation theory does not work. In the gapless phase, we find the logarithmic divergence of the first-order corrections. This leads us to an interesting hypothesis: weak interaction might induce an exponentially small gap in the formerly gapless phase.

\end{abstract}

\maketitle
\thispagestyle{empty}

\section{Introduction}
Superconducting nanostructures and nanodevices are in focus of the condensed matter research community for almost six decades starting from the discovery of Josephson effect\cite{Josephson}. Quantum properties of Josephson-based devices enable sophisticated quantum information technologies\cite{Hybrid,CQE,qubits}. The practical realization of the topological quantum computing paradigm \cite{Kitaev2003} is seen in semiconductor-superconductor nanowire-based nanostructures hosting Majorana states\cite{ReviewM1,ReviewM2}. The superconducting nanostructures vary much in material realization, size, and properties yet can be universally understood in terms of the spectrum of Andreev bound states that depends on phase difference between the superconducting electrodes\cite{QuantumTransport,BeenakkerReview}. There are well-established theoretical tools for analysis and prediction of this spectrum \cite{QuantumTransport,BeenakkerReview}. In this paper, we concentrate on the semiclassical nanostructures with a typical size that is larger then the electron wave length. They involve many transport channels so that their dimensionless (in units of conductance quantum $G_Q\equiv e^2/\pi\hbar$ )conductance $g \gg 1$. The Andreev spectrum is quasi-continuous with a small level spacing $\Delta/g \ll \Delta$, $\Delta$ being the superconducting energy gap.

Most superconducting nanostructures have two terminals like Josephson junctions do. In recent years, there is a considerable increase of interest to multi-terminal superconducting nanostructures, both from theoretical\cite{vanHeck, Padurariu,Yokoyama2015, Weyl, Belzig2020,Repin2019, Meyer2017, Eriksson2017, Levchenko1, Xiaoli, Levchenko2, Levchenko3,Fatemi}  and experimental \cite{Manucharyan2020, Pribiag2020, Finkelstein2019, Courtois, crosswire},  side. Partly, this interest was provoked by the idea that the Andreev levels in N-terminal nanostructures simulate a bandstructure of (N-1)-dimensional material, including its topological properties, and the prediction of quantized transconductance.\cite{Weyl, Eriksson2017} Much research addresses the Weyl points that appear for $N\ge 4$ as topological singularities in the parameter space. \cite{Repin2019, WeylDisks, SpinWeyl, SpintronicsWeyl, Fatemi, Meyer2017}

As to semiclassical nanostructures, it has been discovered that, in distinction from most two-terminal ones, the quasi-continuous spectrum may be gapped or gapless depending on a point on parameter space.\cite{Padurariu} It has been recognized that the gapped phases in a semiclassical structure can be classified with topological numbers.\cite{omega1,omega2} In this case, the presence of gapless phase is readily understood: the domains of the gapless phase should separate the domains of gapped phases of incompatible topology. That has been confirmed experimentally.\cite{omega1,omega2} An extensive investigation of various topologies of this kind is presented in Ref. \cite{Yokoyama2017}. 
 
\begin{figure*}[hbt!]
\begin{center}
	\includegraphics[width=1.8\columnwidth]{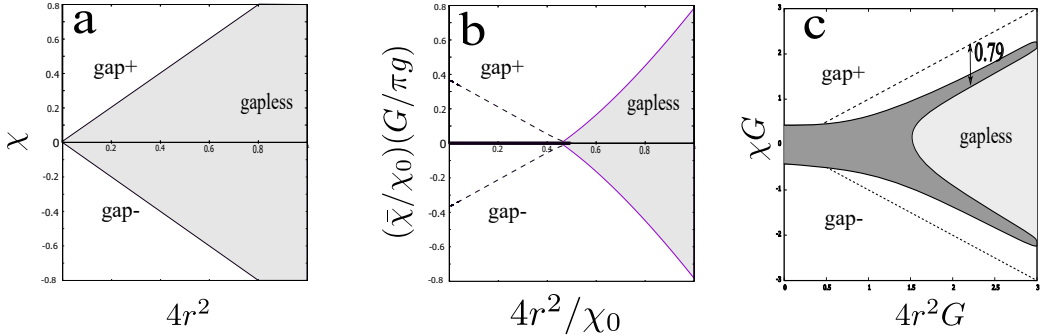} \\
	\caption{ The effect of weak interaction in the vicinity of special point is drastic on a small interaction-defined scale. Diagrams are in parameter space coordinates, $\chi$ is in the direction of main axis, $r$ is the distance from the special point in perpendicular direction. a. No interaction. Two gapped phases of distinct topologies separated by the gapless phase. b. Interaction, semiclassical limit.  First-order transition (thick line) between gapped phases. Dashed lines bound the domain of metastable states. The gappless phase domain is squeezed and shifted from the point. c. Interaction, quantum limit. The gapless phase domain is squeezed, the transition lines are shifted (as indicated by arrow). Dark grey: the domain of strong coupling where the pertubation theory is not applicable.  }
	\label{fig:first}
\end{center}
\end{figure*}

It has been shown that in a wide class of the nanostructures two phases of distinct topology come together in a special point being separated by a domain of gapless phase that becomes infinitesimally thin at the point but yet provides the topological protection.\cite{Xiaoli} (Fig. \ref{fig:first} a).
Each multi-terminal nanostructure can be made effectively two-terminal by setting all terminals to two distinct values of the phase. 
A special point in $N$-dimensional parameter space occurs when these phases are $0$ and $\pi$, that is, the nanostructure is spanned between the opposite values of superconducting order parameter. There are $2^{N-1} -1$ distinct special points in a $N$-terminal nanostructure. It has been also shown \cite{Xiaoli} that for the nanostructures containing tunnel barriers the topological protection may cease so that the domains of topologically distinct phases can come to direct contact: The topological protection is removed in the course of a protection-unprotection transition (PUT). 
In this paper, we concentrate on a close vicinity of a special point and consider the effect of weak interaction. We prove that even a weak interaction provides a drastic effect on the spectrum of Andreev bound states and other characteristics of the nanostructure at a certain scale in parameter space that is determined by interaction and is small if the interaction is weak. We develop an interaction model that encompasses soft confinement and quantum fluctuations in parameter space. Importantly, we derive a universal effective action that provides the adequate description of the situation. The action is compact consisting of four terms only. However, it is hard to analyse involving a non-local order parameter depending on two times, eventually, a matrix in this space. The value of the action is obtained by minimization over this order parameter. 

Owing to this complexity, we are not able to provide in this paper the complete analysis of the action. However, we identify and address two limits: the semi-classical one where the quantum fluctuations do not play a crucial role and the opposite and more interesting limit where the modification of the spectrum is due to quantum fluctuations of the most relevant phase. In the semiclassical limit, we get the exact phase diagram. In the quantum limit, we derive and analyse the first-order quantum correction  that permits to draw qualitative conclusions about the phase diagram and formulate two interesting hypotheses.

Let us already here shortly present the main results obtained (Fig. \ref{fig:first}). The phase diagram without interaction is given in Fig. \ref{fig:first} a. It gives the domains of the gapped and gapless phase in $N$-dimensional parametric space in the vicinity of the point. There is a single axis - main axis - in this space that is orthogonal to the $N-1$-dimensional separation plane between the phases. There is an approximate axial symmetry at the point so $2d$ plot suffucies to present this phase diagram: The coordinate $\xi$is along the main axis  while $r$ gives the distance from the special point in all $N-1$ dimensions orthogonal to the main axis. As promised, we see in the Figure two phases of distinct topology separated by a domain of the gapless phase thinning out at the point. 

The interaction is characterized by a dimensionless conductance $G \gg g$, weaker interaction corresponding to larger $G$. The semiclassical limit of the action holds for 
$G \gg g \gg G/\ln G$. A new exponentially small scale of $\chi$ emerges, $\chi_0 = \exp(-G/g)$. The resulting phase diagram at this scale is presented in Fig. \ref{fig:first} b. We see that the two gapped phases are separated by a first order transition at sufficiently small $r$, and the gap remains finite at the point. The domain of the gapless phase is squeezed and shifted from the point. This implies the absence of topological protection like in the tunnel-junction nanostructures discussed in\cite{Xiaoli}. 

Fig. \ref{fig:first} c presents the results in the quantum limit where $g \ll G/\ln G$. We also see the squeezing of the gapless domain: its boundaries are shifted by the value of $1/G$ in vertical direction. This defines a new small scale of $\chi$. The blacked region in the Figure gives the domain of strong coupling where the perturbation theory does not work: at the boundaries between the gapless and gapped phases and around the special point. We also find that in the gapless phase the first-order correction logarithmically diverges at small energies. 

This inspires us to put forward two hypotheses that should be proved or disproved in the course of further analysis of the strong coupling case. The first hypothesis is that there is no gapless phase and topological protection in the vicinity of the special point: we draw this from continuity with the semiclassical limit. The second hypothesis is motivated by the logarithmic divergence. The divergence may lead to the formation of the interaction-induced exponentially small {\it gap} in the gapless phase. In this way, the gapped-gapless boundaries are crossovers rather than transitions, and the gapless phase is formally an artefact of the non-interacting approximation.

The structure of the paper is as follows. In Section \ref{sec:model} we introduce and  motivate the interaction model in use. We will sketch the derivation of the total action from the quantum circuit theory in Section \ref{sec:derivation}. In Section \ref{sec:action} we will give several representations of the resulting universal action, discuss the scales and define the limits. In Section \ref{sec:nointeraction} we shortly summarize the results for the spectrum near the special point in the absence of interaction. We study the quasiclassical limit in Section \ref{sec:quasi}. More interesting quantum limit is considered in Section \ref{sec:quantum} where we address the first-order interaction corrections. We elaborate on a simplified action that describes the gapped-gapless phase transition in Section \ref{sec:boundary}. In Section \ref{sec:conclude} we formulate hypotheses to be confirmed in the course of future research and finally conclude.

\section{Interaction model}
\label{sec:model}
In superconducting nanostructures, the electromagnetic interaction is usually manifested and described as quantum fluctuations of superconducting phase. \cite{QuantumTransport, Schoen1990}
The scale of the fluctuations is determined by a typical impedance $Z$ of the electromagnetic environment, $\langle(\Delta \phi)^2 \rangle \simeq G_Q Z$. A usual estimation for this impedance is the impedance of free space, with this $\langle(\Delta \phi)^2 \rangle \simeq \alpha$, $\alpha$ being the fine structure constant. This is why the electromagnetic interaction in superconducting nanostructures is usually weak and therefore irrelevant. 

\begin{figure*}[hbt!]
\begin{center}
	\includegraphics[width=0.75\paperwidth]{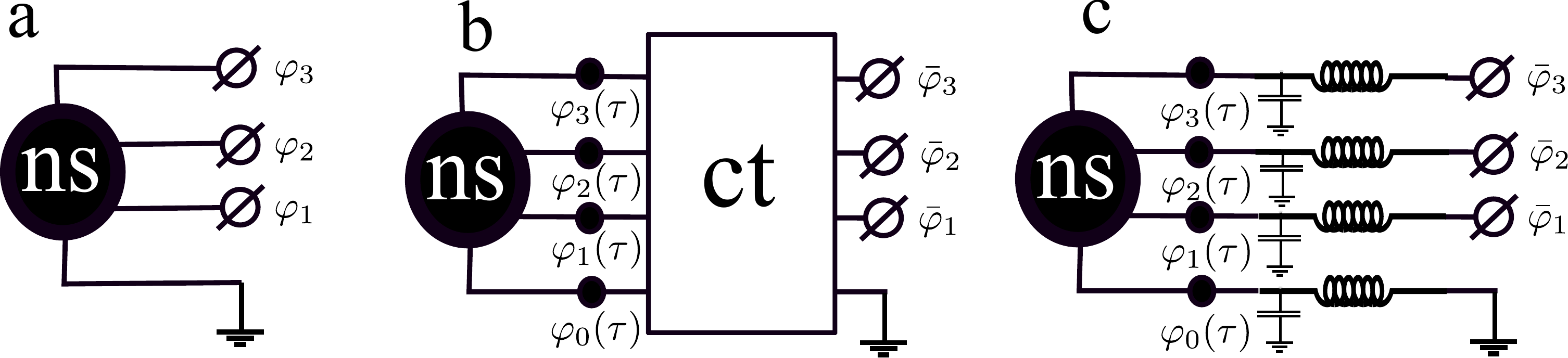} \\
	\caption{Interaction model in use. a. A multi-ternimal nanostructure biased by ideal superconducting terminals with the phases $0,\varphi_{1-3}$. b. Embedding it to a linear circuit makes the terminal phases subject to fluctuations,$\varphi_{0-3}(\tau)$ , and provides the interaction. The circuit confines these quantum variables to $0,\bar{\varphi}_{1-3}$. c. At low freqencies/energies, the curcuit can be presented with inductances and capacitances. Crossinductances(capacitances) are not shown.  }
	\label{fig:circuit}
\end{center}
\end{figure*}

Thus motivated, we set the interaction model by embedding the nanostructure into a linear circuit (Fig. \ref{fig:circuit}). The phases of the superconducting terminals are allowed to fluctuate while the circuit {\it softly confines} these values to time-independent $\bar{\varphi}_i$ that play the role of external parameters set in the experiment.
We employ Matsubara temperature technique (see e.g. \cite{Altland}), so the phases at the terminals are the functions of imaginary time, $\varphi_i(\tau)$, and the partition function is a path integral over these functions ($\hbar =1$ in our system of units),
\begin{align}
Z &= \int \prod_{i,\tau} d \varphi_i (\tau) e^{-{\cal S}}; \\
{\cal S} &= {\cal S}_{\rm ns}(\lbrace\varphi_i(\tau)\rbrace) + {\cal S}_{\rm ct}(\lbrace \varphi_i(\tau) - \bar{\varphi}_i \rbrace);\\
{\cal S}_{\rm ct} &= \frac{1}{2}\int d\tau d\tau' A_{ij}(\tau - \tau') \delta \varphi_i(\tau) \delta \varphi_j(\tau') 
\end{align}
where $A_{ij}(\tau)$ is related to the frequency-dependent admittance of the embedding circuit. The circuit action thus confines the fluctuations $\delta \varphi_i(\tau) \equiv \varphi_i(\tau) -\bar{\varphi}_i$We are interested in low frequencies where the circuit action is readily expressed in terms of the inverse conductance and capacitance matrices of the circuit, 
\begin{align}
{\cal S}_{\rm ct}& = \int d\tau \left(\frac{\hbar^2}{4e^2} (\check{L})^{-1}_{ij} \delta \varphi_i(\tau) \delta \varphi_j(\tau)+ \right.\nonumber\\
& \left. \frac{4e^2}{\hbar^2} C_{ij} \dot{\varphi}_i(\tau) \dot{ \varphi}_j(\tau) \right),
\end{align}
where we denote with "check" the matrices in the space of the terminals.
We have implemented a very similar model to describe interaction effect on Weyl points in superconducting nanostructures.\cite{WeylDisks}

For a generic point in the parameter space one expects a smooth dependence of the nanostructure action on $\varphi_i(\tau)$ so it can be expanded up to the second order in $\delta{\varphi}^i$,
\begin{align}
\label{eq:expansion}
{\cal S}_{\rm ns} &= {\cal S}^{(0)}_{\rm ns} +
+ \frac{\hbar}{2e k_B T} I_i \delta{\varphi}^i + \\&\frac{1}{2}\sum_\omega A^{ij}_{\rm ns} (\omega) \delta{\varphi}_{\omega} \delta{\varphi}_{-\omega}.
\end{align}
The first-order derivatives are proportional to the superconducting currents $I_i$ in the terminals while the second-order terms are related to the frequency-dependent admittance of the nanostructure.
Combining this with the circuit action, we readily obtain two rather trivial and dull corrections to the total action. One is classical and accounts for inductive energy induced by the nanostructure currents in the external circuit, 
\begin{align}
\label{eq:classic}
\delta {\cal S}_{cl} = \frac{1}{2} I_i I_j L_{ij}
\end{align} 
Another one is the renormalization of the nanostructure action by the quantum fluctuations
\begin{align}
\delta {\cal S}_{q} = \frac{1}{2}\sum_\omega A^{ij}_{\rm ns}(\omega) \langle\langle \delta{\varphi}^i_\omega  \delta{\varphi}^j_{-\omega}\rangle\rangle,
\end{align}
with 
\begin{align}
\langle\langle \delta{\varphi}^i_\omega  \delta{\varphi}^j_{-\omega}\rangle\rangle = \frac{4e^2}{\hbar^2}\left[\check{L}^{-1}  + \omega^2 \check{C}\right]^{-1}_{ij}
\end{align}
For estimations, it is instructive to take the superconductive energy scale and represent $\frac{\hbar^2}{4e^2} (\check{L})^{-1} \simeq G \Delta$, 
$G \gg 1$ can be regarded as dimensionless conductance characterizing the external circuit rather than the nanostructure.
With this, the action of the nanostructure is estimated as $g \Delta$, the relative semiclassical correction  as $g/G$, and the relative quantum correction as $1/G$.
The interaction is weak provided $G \gg g \gg 1$.

In a special point, the properties of the nanostructure can be drastically changed by a small variation of the phases and the expansion (\ref{eq:expansion})  does not make sense. In the next Section, we will address the derivation of the appropriate action ${\cal S}_{\rm ns}$ in the vicinity of the special point.

\section{Derivation of the action near a special point}
\label{sec:derivation}
In general, the nanostructure action is computed from the Nambu-structured electron Green functions $\hat{G}(\tau,\tau'; {\bf r},{\bf r})$ defined in each point ${\bf r}$ of the structure that are subject to time-dependent superconducting order parameters $\Delta_i e^{i \varphi_i(t)}$ in the adjacent superconducting leads. We will use the method of quantum circuit theory \cite{QuantumTransport} which is a finite-element approximation to the actual coordinate-dependent Green functions  that is suitable for semiclassical nanostructures. The line of derivation  is similar to that of \cite{Xiaoli} yet it is adjusted to time-dependent fields.

\subsection{Quantum circuit theory}
\label{sec:qct}
In quantum circuit theory, the nanostructure is represented as a set of nodes connected by connectors: the scatterers characterized by a set of transmission coefficients $T_p$. In each node, the Green function is represented as a matrix $\hat{G}$ that incorporates Nambu structure and two time indices. The matrix satisfies 
\begin{align}
  \hat G^2 = \hat 1 && \Tr \hat G = 0
\end{align}

The total action is a sum of contributions of the connectors. A contribution of a connector is expressed in terms of the Green functions at its ends, $\hat{G}_{A,B}$ 
\begin{equation}
  S = \frac{1}{2} \sum_p \Tr \left\{
    \log \left[
      1 + \frac{T_p}{4} \left( \hat G_A \hat G_B + \hat G_B \hat G_A - 2 \right)
    \right]
  \right\}. 
\end{equation}
Here, the trace incorporates imaginary time, $\Tr[A] =  \int_0^\beta \Tr_{\rm Nambu}[A(\tau, \tau)] d \tau$, with $\beta = 1/k_B T$.

It is convenient to incorporate the information about the transmission distribution to the characteristic function of a connector $\mathcal{F}(x)$ defined as:
  \begin{equation}
    \mathcal{F}(x) = \sum_p \log \left[
      1 - \frac{T_p}{2} (1 - x)
    \right]
  \end{equation}
where the sum is over all transmission eigenvalues. For a tunnel junction $\mathcal{F}_T(x) = - ({G_T}/{2 G_Q})(1 - x)$, for a ballistic contact $\mathcal{F}_B(x) = ({G_B}/{G_Q}) \log \left[ \frac{1 + x}{2} \right]$ and for diffusive connector $\mathcal{F}_D(x) = ({G_D}/{8 G_Q}) \arccos(x)^2$. 

With this, a connector action reads
\begin{equation}
  S = \frac{1}{2} \Tr \left\{
    \mathcal{F}((\hat G_A \hat G_B + \hat G_B \hat G_A)/2)
  \right\}.  \label{eq:2}
\end{equation}

 A subset of nodes are terminals where the Green functions are fixed to ($\eta_z$ is a Pauli matrix in Nambu space)
\begin{equation}
  \hat{G}_i(\tau,\tau') = e^{-i \eta_z \varphi_i(\tau)/2} G^{(0)}_i(t - t') e^{i \eta_z \varphi_i(\tau')/2}
\end{equation}
with $G^{(0)}_i$ to be given in energy representation as
\begin{equation}
  G^{(0)}_i(\varepsilon) = \frac{1}{\sqrt{\Delta_i^2 + \varepsilon^2}} \left[
    \begin{array}{cc}
      \varepsilon & \Delta_i  \\
       \Delta_i & -\varepsilon
    \end{array}
  \right]
\end{equation}

Importantly, the overall action has to be minimized with respect to $\hat{G}$ in all nodes. The result of the minimization will give the actual ${\cal S}_{\rm ns}(\lbrace \varphi_i(\tau)\rbrace)$.

\begin{figure*}[hbt!]
\begin{center}
	\includegraphics[width=0.75\paperwidth]{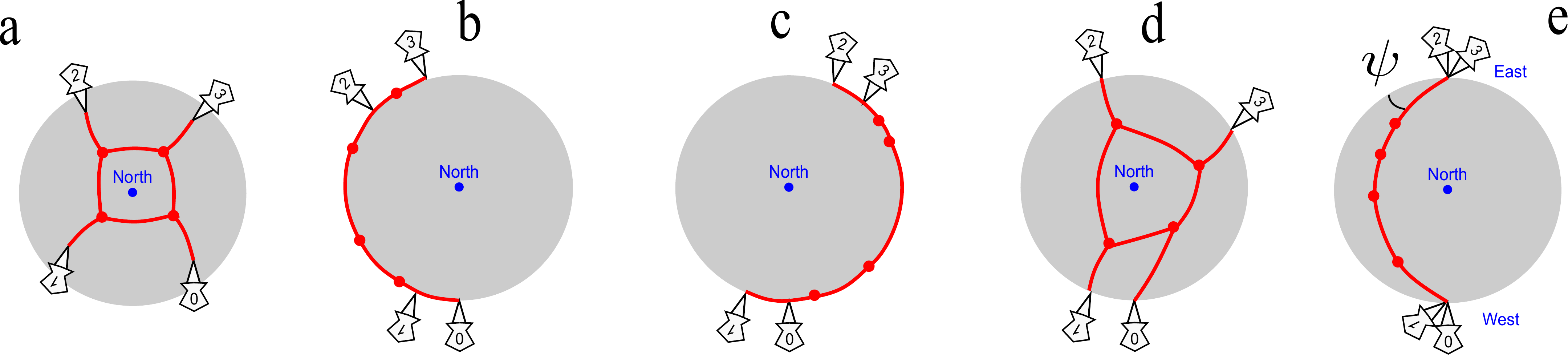} \\
	\caption{Rubber thread representation of quantum circuit theory. a. Four-terminal, four-node, eight-connector network. Finite $\varepsilon$, the terminals are above the equator. b. Zero energy, terminals are at the equator, the network spanned along the equator. Gapped phase. c. Gapped phase of different topology.
	d. Gapped phase: the nodes are spanned over the hemisphere, while the terminals are pinned at the equator.
	e. Special point. The terminals are located at opposite points of the sphere (West and East pole). The network is spanned along the arc, the action is degenerate with respect to $\psi$.} \label{fig:rubber}
	\end{center}

\end{figure*}

\subsection{Rubber thread representation}
\label{sec:rtr}
Let us here recall the representation of the quantum circuit theory that looks naive but is in fact very instructive energizing common intuition. In this subsection, we restrict ourselves to time-independent $\varphi_i$. 

In this case, the Green functions in the nodes can be minimized separately at each energy and are $2 \times 2$ Hermitian matrices to be represented with real vectors $\hat{G}(\varepsilon)\to \vec{g} \cdot \vec{\eta}$. The vectors are associated with points on a sphere, since $\hat{G}^2=1$ implies $\vec{g}^2=1$. Eventually, $\vec{g}$ are either in upper or lower hemisphere depending on the sign of $\varepsilon$.

The connectors are associated with rubber threads connecting the nodes, and the action with the elastic energy of the threads that depends on the angle between the vectors $\vec{g}_{A,B}$ at the ends of a thread. The elastic energy tries to bring all nodes to one point. However, the vectors $\vec{g}$ are fixed in the terminals, and the whole nanostructure is associated with a rubber thread network pinned at the terminals  and spanned over the hemisphere to minimize the elastic energy (Fig. \ref{fig:rubber} a). 

The case $\varepsilon =0$ is special. In this case, the pins are exactly at the equator. The $z$-coordinate of a node is associated with the density of states at this node, the superconductivity vanishing at the North pole. Depending on the positions of the pins, the network can be either spanned over the equator (gapped phases, 
Fig. \ref{fig:rubber} b and c) or over the whole hemisphere (gapless phase, Fig. \ref{fig:rubber} d).

\subsection{The special symmetry at a special point}

In general, the minimization of the action unambiguously determines $\hat{G}$ as functions of the terminal $\hat{G}_i$. However, there is an ambiguity precisely at a special point and zero energy. The rubber thread representation helps to understand why. In a special point, all pins are at precisely opposite positions at the equator: let us call them West and East poles. The network is spanned over an arc connecting the poles. Elastic energy does not fix the position of the arc given by the angle $\psi$ (Fig. \ref{fig:rubber} e). There is an extra rotational symmetry about the axis that does through the poles.

We fix the Green functions of the terminals to  the $y$ axis,  $\hat{G}_i =\zeta_i \eta_y$ where $\zeta_i = \pm 1$ gives if a given terminal is at the West or East pole. We parametrize  the ambiguous Green functions in the nodes as follows: 
\begin{equation}
\vec{g}_a = (- \sin \theta_a \cos \hat \psi, \cos \theta_a, - \sin \theta_a \sin \psi).
\end{equation}
Here and in the following, we index the terminals with $i,k,l...$ and the nodes with the letters from the beginning of the alphabet: $a,b,c$.
The total action is a sum over connectors and does not depend on $\psi$,
\begin{align}
S = \sum_{i,a} {\cal F}_{i a}(\zeta_i \cos \theta_a) + \sum_{a>b} {\cal F}_{ab}(\cos(\theta_a - \theta_b))
\end{align}
while $\theta_a$ are specific for a nanostructure and are determined from the minimization of above action. Here, ${\cal F}_{ab}$ refers to a connector connecting the nodes $a$ and $b$,${\cal F}_{ia}$ to a connector connecting node $a$ and a terminal $i$.

Now we come to an important step: in the limit of low energies, that is relevant near a special point, the rotational angle is not a number, it can be an arbitraty matrix in two times, $\phi \to \hat{\phi} \equiv \psi(\tau, \tau')$. This matrix parametrizes the whole set of degenerate solutions for time-dependent Green functions at the special point:   
  \begin{equation}
  \hat G_a = e^{i \eta_y \hat \psi/2} \left[
    \begin{array}{cc}
      0 & i e^{i \theta_a} \\
      -i e^{-i \theta_a} & 0 
    \end{array}
  \right] e^{-i \eta_y \hat \psi/2}
\end{equation}

This high degeneracy, either for a number $\psi$ or for the whole matrix $\hat{psi}$ implies that in order to describe the situation in the vicinity of a special point, we have to consider the terms that break the special symmetry described and actually fix $\hat{\psi}$. 
This $\hat{\psi}$ can be regarded a specific order parameter for  the vicinity of the special point.

\subsection{Symmetry-breaking terms}
As discussed in REF, there are actually three distinct terms of this sort. The first term corresponds to deviations of the terminals from the equator at non-zero energy: energy term. The second one describes the first-order corrections owing to the shift of the terminals along the equator: the shift of the phases in $N$-dimensional parameter space from the special point. Since it is linear in phase deviations, it is sensitive to the shift along a single direction in the N-dimensional parameter space, that we call the main axis. The third term describes the influence of the shift in the direction perpendicular to the main axis: it has to be the second-order term.

To compute the first term, we expand the Green functions of the terminals as
\begin{equation}
\hat G_T = \zeta_T \eta_y + \frac{\varepsilon}{\Delta}_i \eta_z
\end{equation}
and collect first-order corrections induced in the connectors adjacent to the nodes. This gives 
\begin{align}
{\cal S}_1 =  -  \tilde{g}  \Tr \left[
      \hat {\epsilon} \sin \hat {\psi}
    \right]\\
\tilde{g} = \sum_{i,a} \frac{1}{\Delta_i} \mathcal{F}'_{i a}(\cos \theta_a) \sin \theta_a
\end{align}
The constant $\tilde{g}$ has a meaning of the maximum inverse level splitting of  Andreev states. The above expression is for the case of short nanostructure $\Delta \ll E_{\rm Th}$, $E_{\rm Th}$ being the Thouless energy. In general case, we need to take into account at the circuit theory level the so-called "leakage" terminals \cite{QuantumTransport} that account for finite volume of the nodes. This gives an addition to $\tilde{g}$ in terms of inverse level spacings $\delta^a_S$ in all nodes,
\begin{equation}
\tilde{g} \to \tilde{g} + \sum_{a} \frac{1}{\delta^a}.
\end{equation}
Thus in the opposite limit $\Delta \gg E_{\rm Th}$ $\tilde{g}$ is the inverse level spacing for normal electron states in the whole structure.

Next, we consider the effect of phase deviations $\chi_i(\tau)$ from the special point, $\varphi_i(\tau) = \pi/2 + \pi \zeta_i + \chi_i(\tau)$ The expansion of the terminal Green functions read
\begin{equation}
\hat{G}_i = \zeta_i \eta_y  - \zeta_i \hat{\chi}_i \eta_x
\end{equation}
where $\hat{\chi}_i \equiv \chi(\tau) \delta(\tau - \tau')$. Collecting the first-order corrections to the action of adjacent connectors gives
\begin{align}
{\cal S}_2 &= \sum_{i} F_i \Tr \left[ \hat \chi_i \cos \hat \psi \right];
  \label{eq:s2}\\
  F_i &=  \zeta_i \sum_{a} \mathcal{F}'_{i a}(\cos \theta_a) \sin \theta_a
\end{align}
$F_i$ is a dimensionless vector in the space of the terminals with the amplitude proportional to the dimensionless conductance $g$ of the structure. 

The computation of the third term is more involved since it concerns the second order corrections. It is contributed by the first-order terms coming from the second-order deviations of the terminal Green functions and the quadratic reaction of the network on the first-order deviations of those. To find the latter, we expand the Green functions in all nodes till quadratic terms in their deviations $\hat{w}$, find the terms coupling these deviations and first-order deviations of the terminal Green functions, and minimize with respect to $w$. The result reads ($\hat{U} \equiv e^{i \hat{\psi}}$)
\begin{align}
{\cal S}_3 = \frac{1}{2} \sum_{i,j} H_{ij} \Tr[\hat{U} \hat{\chi}_i \hat{U} \hat {\chi}_j + \hat{U}^{-1} \hat{\chi}_i \hat{U}^{-1} \hat{\chi}_j]
\end{align}
$\check{H}$ being the matrix in the space of the terminals. It is dimensionless and also scales as the dimensionless conductance $g$ of the nanostructure. Eventually, the vector $\chi_i$ in this expression has to be orthogonal to $F_i$, since the second corrections in these direction are negligible in comparison with the first order term taken into account in Eq. \ref{eq:s2}.

Its concrete expression is rather clumsy, 
\begin{align}
\check{H} = 2 + \frac{1}{2} \check{B}^{T} \check{Q}^{-1} \check{B} + \frac{1}{2} \check{A}^{T} \check{P}^{-1} \check{A};
\end{align}
Where the matrices $\check{A}$, $\check{B}$
connect the terminals and nodes,
\begin{align}
A_{ai} &= \frac{1}{2} (\mathcal{F}''_{ai} \sin^2 \theta_a + \zeta_i \mathcal{F}'_{ai} \cos \theta_a),\\
\check{B}_{ai} &=\frac{1}{2} \zeta_i \mathcal{F}'_{ai} \cos (\theta_a) 
\end{align}
and  $\check{Q},\check{P}$ are matrices in the space of the nodes,
\begin{align}
P_{ab} &= \frac{1}{2} \delta_{ab} \left[- \sum_i A_{ai} - 
      \sum_c ( \mathcal{F}'_{ac}\cos(\theta_a - \theta_c) \right. \nonumber\\
&\left.	  -\mathcal{F}''_{ac}  \sin^2(\theta_a - \theta_c) ) \right] \nonumber\\
 &- \mathcal{F}'_{ab} \cos (\theta_a - \theta_b) +
\mathcal{F}''_{ab} \sin^2 (\theta_a - \theta_b)\\
Q_{ab} &= - \delta_{ab}( \sum_i B_{ai} + \nonumber\\
 &\frac{1}{2}  \sum_c \mathcal{F}'_{ac} \cos(\theta_a - \theta_c)) + \mathcal{F}'_{ab}.
\end{align}
In all above expressions,
$\mathcal{F}'_{ab} \equiv \mathcal{F}'_{ab}(\cos(\theta_a-\theta_b)$, $\mathcal{F}'_{ai} \equiv \mathcal{F}'_{ab}(\cos(\theta_a)$ and similar for $\mathcal{F}''$. 

The actual $\hat{\phi}$ as a functional of $\chi(\tau)$ is found by minimization of all three symmetry-breaking terms. Therefore the answer for the relevant part of the nanostructure action reads 
\begin{equation}
{\cal S}_{\rm ns} = \mathop{{\rm min}}_{\hat{\psi}} [ {\cal S}_1 +{\cal S}_2 +{\cal S}_3 ].
\end{equation}

\section{The action}
\label{sec:action}
In this Section, we will present the resulting action in several equivalent forms, discuss the energy and parametric space distance scales, and establish the relevant simpler limits.
We collect the results of the previous Sections into the following form:
\begin{align}
{\cal S}& = \mathop{{\rm min}}_{\hat{\psi}} {\rm Tr}[ - \tilde{g} \hat{\varepsilon}  \sin \hat{\phi} + F_i \hat{\chi}_i \cos \hat{\psi} + \nonumber\\
&\frac{1}{2} H_{ij} (\hat{U} \hat{\chi}_i \hat{U} \hat {\chi}_j + \hat{U}^{-1} \hat{\chi}_i \hat{U}^{-1} \hat{\chi}_j)] +\nonumber \\
& \int d\tau \left(\frac{\hbar^2}{4e^2} (\hat{L})^{-1}_{ij} \delta \chi_i(\tau) \delta \chi_j(\tau)+ \right. \nonumber\\
& \left. \frac{4e^2}{\hbar^2} C_{ij} \dot{\chi}_i(\tau) \dot{ \chi}_j(\tau) \right),
\end{align} 
$\delta{\chi} \equiv \chi(\tau) - \bar{\chi}$. Let us do the following rescalings and coordinate changes. First of all, we make the energy dimensionless measuring it in units of $\Delta$: $\epsilon = \varepsilon/\Delta$. Here $\Delta$ is the superconducting energy gap if it is the same in all leads or the maximum of $\Delta_i$. Its precise value is not important since near the special point the relevant energy scale is much smaller than $\Delta$. We define $g \equiv \tilde{g}\Delta$ as the measure for dimensionless conductance of the nanostructure. Next, we change the coordinates in the phase parametric space. The coordinate in the direction of the main axis is defined as $\chi = F_i\chi_i/g$, $\chi$ being dimensionless and small as far as we are in close vicinity of the special point.
We project the matrix $H_{ij}/g$ into $N-2$ directions orthogonal to the main axis, diagonalize it and introduce the dimensionless coordinates $r_k = \sqrt{H}_k h^{(k)}_j \chi_j$, $H_k$, $h^{(k)}_j$ being the eigenvalues and corresponding eigenstates of this matrix. We disregard the capacitance terms in the circuit action assuming that the frequency scale of the relevant quantum fluctuations is much smaller than $1/\sqrt{LC}$. With this, we rewrite the action as follows:
\begin{align}
{\cal S}&= -\frac{g}{2} \mathop{{\rm min}}_{\hat{\psi}}[{\rm Tr}[ \hat{U}^\dagger \hat{A} - \hat{U}\hat{r}_k \hat{U} \hat{r}_k + h.c.
]] \nonumber \\
&+\frac{G}{2} {\rm Tr} [ (\delta \hat{\chi}^2) + \frac{G^k}{G} \delta \hat{\chi} \delta\hat {r}_k + \frac{G^{kl}}{G} \delta \hat{r_k} \delta \hat{r}_l] 
\label{eq:actionwithg}
\end{align} 
with
\begin{equation}
A,A^{\dagger} = \hat{\chi} \pm i \epsilon.
\end{equation} 
Here, we rewrote the inverse inductance matrix in new coordinates in the parameter space. The nanostructure is characterized by dimensionless conductance $g$, and the circuit by the dimensionless conductance $G \gg g$, the larger value $G$ corresponding to smaller interaction. Two dimensionless energy scales are defined by $|\bar{\chi}|$, $\sum_k\bar{r}^2_k \equiv r^2$. Without interactions, $|\chi| > 4 r^2$ corresponds to the gapped phase, $|\chi|< 4 r^2$ to the gapless one. 
 

It may seem that the most pronounced interaction effect comes from the fluctuations of $r_k$. Naively, the coefficient $r^2$ in front of the third term would be replaced by $\langle \langle r^2 \rangle \rangle$ and would remain finite even at $\bar{r}=0$ resulting in a finite width of the separating gapless phase domain. More careful analysis shows that this does not happen. Owing to the ordering of the operators $\hat{r}$ and $\hat{U}$ the fluctuations of $r$ eventually lead for insignificant corrections $\simeq G^{-1}$ to the second term. This inspires us to disregard the fluctuations of $\hat{r}$. Indeed, except the crucial point mentioned, their fluctuations should provide lesser effect than those of $\chi$ that enter the action in the first order. So in further analysis, we disregard the fluctuations of $r_k$ skipping the last two terms in the action (\ref{eq:actionwithg}).

With these assumptions, let us look at the limits. To start with this, let us assume no fluctuations of $\chi$ as well and replace it with a  time-independent value. It may seem that this would lead to a trivial correction like the one given by Eq. \ref{eq:classic} and would not modify the spectrum significantly. However, the inverse inductance of the nanostructure logarithmically diverges at $r^2 \to 0$, $\partial {E}/\partial {\chi} = -(g/\pi)\chi  \ln(1/|\chi|)$ so it successfully competes with the formally larger confining term at an exponentially small scale of $\chi$, $\chi_0 \equiv \exp( - \pi G/g)$. This scale defines an interesting quasi-classical limit detailed in Section \ref{sec:quasi}.

Let us understand the significance of quantum fluctuations in this limit. The quantum fluctuation of $\chi$ can be generally estimated as 
\begin{align}
\label{eq:fluctuationsestimate}
\langle\langle \chi^2 \rangle \rangle \simeq \frac{\epsilon_s}{G},
\end{align}
$\epsilon_s$ being the relevant frequency scale.
In semi-classical regime this scale is defined by $\chi_0$. Comparing the fluctuation and $\chi_0$ itself, $\langle\langle \chi^2 \rangle \rangle \leftrightarrow \chi_0^2$ we obtain that the quantum fluctuations can be neglected if $\chi_0 \gg 1/G$, that is, if $g \gg G/\pi\ln G$. Since $g \ll G $, this sets a rather narrow but relevant range of $g$.

If $g \ll G/\pi \ln G$, the quantum fluctuations destroy the logarithmic divergence and the nanostructure contribution to the action does not compete with that of the circuit, ${\cal S}_{\rm ns} \ll {\cal S}_{\rm ct}$. The way to proceed in this case is to expand $e^{-{\cal S}}$ in terms of ${\cal S}_{\rm ns}$ keeping the first term of the expansion. The nanostructure is thus characterized by  ${\cal S}_{\rm ns}$ averaged over Gaussian quantum fluctuations produced by the circuit,
\begin{align}
\label{eq:averagedaction}
\overline{{\cal S}_{\rm ns}}&= \int \prod_\tau d\chi_\tau {\cal S}_{\rm ns}(\lbrace \chi(\tau)\rbrace) \nonumber \\
&\times \exp\left(- \frac{G}{2} \int d \tau (\chi(\tau) -\bar{\chi})^2 \right)
\end{align} 
We compute the first-order ($\propto G^{-1}$) interaction correction in Section \ref{sec:quantum}. To estimate where it becomes significant, we compare $\langle\langle \chi^2 \rangle \rangle$ with $\chi$ taking $\chi$ as the relevant frequency scale. With this, the interaction leads to significant modification of the spectrum at small scale $\chi \simeq 1/G$. (Fig. \ref{fig:first} c)

If we depart from the special point in the orthogonal direction at distances $r^2 \gg 1/G$, the interaction is significant in a narrow strip at the boundary between the gapped and gapless phase, that is shifted by $\simeq G$, $\chi_c = 4 r^2 - 0.79/G$ (as computed in the next Section). To estimate the width of strip $\delta \chi_c$, we compare it with the quantum fluctuation (Eq. \ref{eq:fluctuationsestimate}) taking into account that the relevant frequency scale is defined by $\delta \chi_c$ itself, $\epsilon_s \simeq \chi_c (\delta \chi_c /\chi_c)^{3/2}$ (see Section \ref{sec:boundary} for details). This 
gives $\delta \chi_c \simeq G^{-1} (\chi_c G)^{-1}$. (Fig. \ref{fig:first} c)

Let us complete this Section with giving several equivalent forms of the action that are convenient for concrete calculations. If we neglect the fluctuations of $r_k$, we can rescale the action measuring frequency in units of $4 r^2$ and introducing rescaled $X \equiv \chi/(4 r^2)$. In this form, the action reads:
\begin{align}
\label{eq:actionrescaled1}
\cal{S} &=-\frac{4 r^2 g}{2} \mathop{{\rm min}}_{\hat{\psi}}\left[{\rm Tr}\left[ \hat{U}^\dagger \hat{A} - \frac{\hat{U}^2}{4} + h.c\right]\right] \nonumber\\+& \frac{1}{2 \lambda} \rm Tr[(\hat{X} -\bar{X})] 
\end{align}
with $\lambda = (4r^2 G)^{-1}$, $\hat{A} = \hat{X} + i \hat{\epsilon}$. 

Apart from an insignificant constant, the third term containing two $\hat{U}$ matrices  can be presented as a minimum over an additional Hermitian operator $\hat{p}$,
\begin{align} 
\frac{1}{2} {\rm Tr} [\hat{U}\hat{r}_k \hat{U} \hat{r}_k+ \hat{U}^\dagger\hat{r}_k \hat{U}^\dagger \hat{r}_k ]=\\ 
\mathop{{\rm min}}_{\hat{p}_k} {\rm Tr} [ \frac{\hat{p}^2}{2} + i\hat{p}(\hat{U}\hat{r}_k - \hat{r}_k \hat{U}^\dagger)]
\end{align} 
This suggests the following form of the action:
\begin{align}
{\cal S} =& -\frac{g}{2} \mathop{{\rm min}}_{\hat{\phi},\hat{p}_k}[{\rm Tr}[ \hat{U}^\dagger \hat{A} - \sum_k\hat{p}_k^2/2 + h.c.]\\+
&\frac{G}{2} {\rm Tr}[(\hat{\chi}-\bar{\chi})^2]
\end{align}
with $A = \hat{\chi} - i 2\hat{p}_k\hat{r}_k + i \hat{\epsilon}$. 

The latter trick can be also applied to the rescaled form of the action (Eq. \ref{eq:actionrescaled1}). In this case, we may use a single auxiliary operator $\hat{p}$ that can be regarded as an addition to $\hat{\epsilon}$. So it is convenient to rewrite the action as 
\begin{align}
\label{eq:actionrescaled2}
\cal{S} &=-\frac{4 r^2 g}{2} \mathop{{\rm min}}_{\hat{\psi,\tilde{\epsilon}}}[{\rm Tr}[ \hat{U}^\dagger \hat{A} +\hat{A}^\dagger\hat{U} - (\hat{\epsilon}-\hat{\tilde{\epsilon}})^2 ] \nonumber\\+& \frac{1}{2 \lambda} \rm Tr[(\hat{X} -\bar{X})] 
\end{align}
with $A = \hat{X} + i \hat{\tilde{\epsilon}}.$

\section{Non-interacting spectrum}
\label{sec:nointeraction}

\begin{figure}[hbt!]
\begin{center}
	\includegraphics[width=0.8\columnwidth]{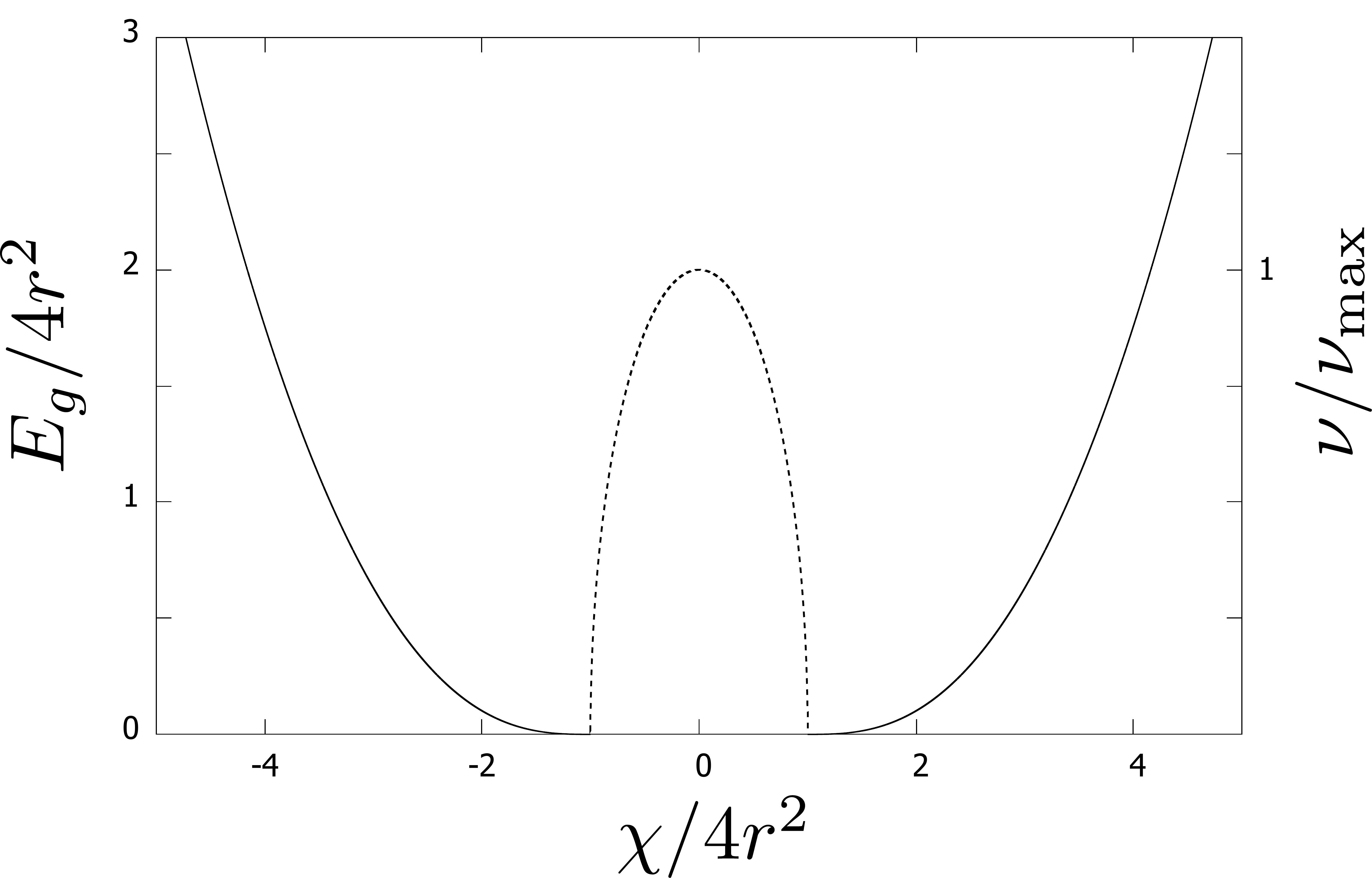} \\
	\caption{The spectrum of the nanostructure without interaction. We plot the gap in the gapped phases (solid line) and the density of states in the gapped phase $|\chi|<4r^2$}
	\label{fig:noninteracting}
	\end{center}	
\end{figure}
Let us reproduce the results without interaction obtained in \cite{Xiaoli} and extend those for the needs of the present paper. In this case, we can replace fluctuating $\hat{\chi}$ with a constant $\chi$ regarding it as a parameter. We will use the rescaled action given by Eq. \ref{eq:actionrescaled2}. All operators involved are diagonal in energy representation so we replace them by the corresponding eigenvalues. The minimization with respect to $\hat{\psi}$ gives
\begin{align}
U \equiv e^{i\psi} = \frac{X + i \tilde{\epsilon}}{\sqrt{\tilde{\epsilon}^2 + X^2}}
\end{align}
and that with respect to $\tilde{\epsilon}$ gives $\tilde{\epsilon}(\epsilon)$ in the following implicit form:
\begin{align}
\label{eq:forepsilontilde}
\epsilon = \tilde{\epsilon}\left(1-\frac{1}{\sqrt{X^2 + \tilde{\epsilon}^2}}\right).
\end{align}
Let us first solve these equations at zero energy. If $|X|>1$, the only solution of Eq. \ref{eq:forepsilontilde} is $\tilde{\epsilon}(0)=0$. With this, $U(0)={\rm sgn}(X)$, $\psi = \pi/2(1-{\rm sgn}(X))$ and the density of states $\simeq \sin \psi$ is zero. We are in a gapped phase, topologically distinct phases being realized at positive/negative $X$. If $|X|<1$, another solution is realized,  $\epsilon(0) = \sqrt{1-X^2}$.
This gives a non-zero density of states at zero energy :
\begin{align}
\frac{\nu}{\nu_{max}} = \sqrt{1-X^2}.
\end{align}
We are in the gapless phase separating the gapped ones. 

To find the gap in the gapped phases, we look at the solutions at imaginary $\epsilon$ and find the root of $\partial \epsilon/\partial \tilde{\epsilon}=0$. This gives 
\begin{align}
E_g = \left(|X|^{2/3} -1\right)^{3/2},
\end{align}
so the gap closes at $|X|=1$ and approaches $|X|$ at $|X| \gg 1$. We plot these results in Fig. \ref{fig:noninteracting}. 

We can also compute the energy of the nanostructure. For our purposes, we only need its derivative with respect to $X$, that is given by 
\begin{align}
\frac{\partial E}{\partial X} &= - {4 r^2 g} \int \frac{d \epsilon}{2\pi} \cos \psi \nonumber \\=& -{4 r^2 g} \frac{d \epsilon}{2\pi} \frac{X}{\sqrt{X^2 +\tilde{\epsilon}}^2}.
\end{align} 
This integral can be easily evaluated by transforming the integration variable with the help of Eq. \ref{eq:forepsilontilde}. Note the logarithmic divergence at $\epsilon  \to \infty$, that we cut at $ |\epsilon| \approx |\tilde{\epsilon}| =\omega_D \gg |X|$. With this, the result in the gapped phase reads 
\begin{align}
\frac{\partial E}{\partial X} &= - \frac{4 r^2 g}{2\pi} X\left(\ln \left(\frac{2 \omega_D}{|X|}\right)+Z(X)\right);\label{eq:der} \\
Z(X) & \equiv - \frac{\pi}{4|X|} +\frac{\Theta(1-|X|)}{2} \left(\sqrt{1-X^2} \right. \nonumber \\
&+ \left. X\arctan\left(\frac{\sqrt{1-X^2}}{X}\right)\right) \label{eq:zX} 
\end{align}
Thus, in the gapless phase, $Z(X)$ acquires an addition given by the second term in Eq. \ref{eq:zX}.

\section{Quasiclassical limit}
\label{sec:quasi}
\begin{figure}[hbt!]
\begin{center}
	\includegraphics[width=0.8\columnwidth]{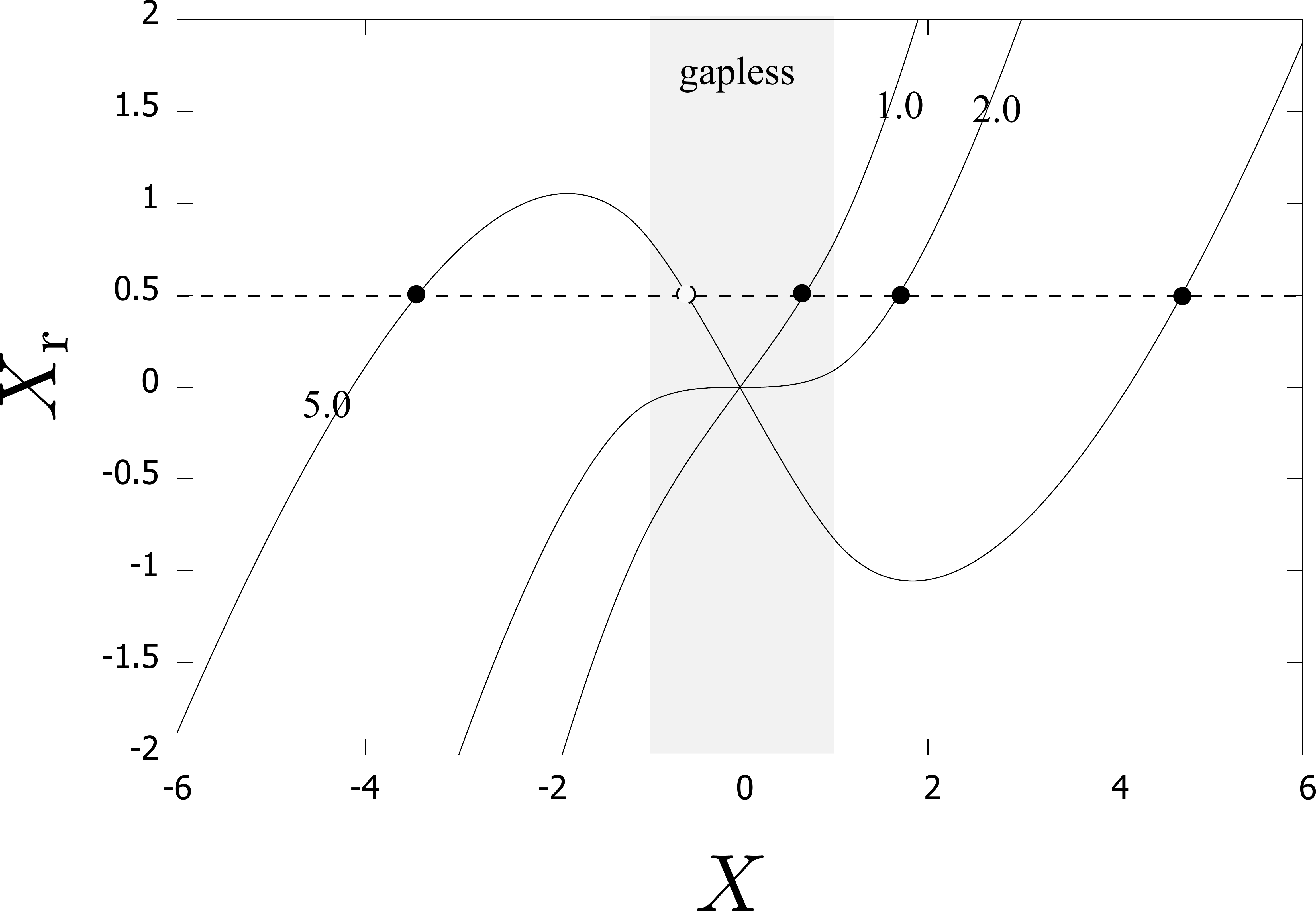} \\
	\caption{Quasiclassical limit. Solving the self-consistency Eq. \ref{eq:chi0rescaled}. The curves $X_{\rm r}(X)$ resembling van der Waals isotherms are plotted for several values of $\chi_0$ given in the curve labels. Stable solutions for $X_{\rm r}=0.5$ are given by black dots. The solutions within the grey strip correspond to gapless phase. }\label{fig:vanderwaals}
	\end{center}	
\end{figure}
In this Section, we concentrate on the quasiclassical limit of the interacting problem. As outlined in Section \ref{sec:action}, it is realized at $G \gg g \gg G/\ln G$. We can disregard the fluctuations of $\chi(\tau)$ treating it as a time-independent variable $\chi$. However, in a close vicinity of the special point the external circuit fails to confine $\chi$ to the parameter $\bar{\chi}$, as expected in the non-interacting limit $G \gg 1$. The reason for this is the logarithmic divergence of the inverse inductance of the nanostructure that was just quantified in the previous Section. The actual value of $\chi$ has to be determined from the minimization of the action given by Eq. \ref{eq:actionwithg} or Eq. \ref{eq:actionrescaled2},
\begin{align}
\frac{\partial E(\chi)}{\partial\chi} + G(\chi - \bar{\chi})=0.
\end{align}
We will use the rescaled action. With Eq. \ref{eq:der}, the above self-consistency equation reads
\begin{align}
X(\ln \left(\frac{2\omega_D}{|X|} + Z(X))\right)= \frac{G}{\pi g} (X-\bar{X})
\end{align}
Actually, it resembles a well-known BCS self-consistency equation (see, e.g. \cite{gen66}) for the superconducting order parameter $\Delta$ , $X$ playing the role of $\Delta$, that relies on a similar logarithmic divergence of a response function. We will get rid of the explicit cut-off $\omega_D$ by very same method as in BSC theory: 
We substute $(G/\pi g)$ with $\ln(2\omega_D/\chi_0)$, which defines new exponentially small scale
\begin{align}
\chi_0 = 2 \omega_D \exp\left(- \frac{\pi G}{g}\right),
\end{align}
and also rescale the external parameter $X_{\rm r} =\bar{X} (G/\pi g) $. With this, the self-consistency equation becomes an  expression for $X_{\rm r}$ in terms of $X$, 
\begin{align}
\label{eq:chi0rescaled}
X_{\rm r} = -X(\ln(\chi_0/X) + Z(X)). 
\end{align}
This is the rescaled equation so $r^2 \to 0$ corresponds to $\chi_0 \gg 1$. In this limit, one can neglect $Z(X)$. At $X_{\rm r}=0$, that is, precisely at the special point we encounter two solutions $X = \pm \chi_0$ corresponding to two gapped phases with the gap $\chi_0$. These two solutions coexist up to $|X_{\rm}| = \chi_0/e$. Their energies differ except at $X_{\rm r}=0$, where we encounter the line of first order transition (Fig. \ref{fig:first}. a). For further qualitative analysis, it is instructive to plot  Eq. \ref{eq:chi0rescaled} for several values of $\chi_0$(Fig. \ref{fig:vanderwaals})where the resemblence to van der Waals isothermes becomes apparent. We see that the line of constant $X_{\rm r}$ gives either one (at $\chi_0 <2$) or three solutions (at $\chi_0 >2$) for $X$, two of them being stable. The first-order transition line thus ends at $\chi_0 =2$ (This corresponds to the critical distance $4 r^2 = 0.5 \chi_0$ in non-rescaled units). The lines where the stable solutions dissapear (dashed lines in Fig. \ref{fig:first} b) are defined from the positions of the extrema of the curves drawn in Fig. \ref{fig:vanderwaals} and were obtained by the implicit plot
\begin{align}
X_{\rm r} &= X^2 Z'(X) -1 \\
\ln \chi_0 &= \ln|X| - Z(X) -X Z'(X) +1
\end{align} 

The solutions with $|X|<1$ correspond to the gapless phase. Substituting $X=\pm 1$ to Eq. \ref{eq:chi0rescaled} gives the lines of phase transitions between the gapless and gapped phases $X_{r} = \pm (\pi/4 - \ln \chi_0)$. We note that the gap disappears already at the first order transition line, at $\chi_0 = \exp(\pi/4) \approx 2.19$ (in non-rescaled coordinates, this corresponds to $4 r^2 = 0.46 \chi_0$.

The full phase diagramm is presented in Fig. \ref{fig:first} b in coordinates $4 r^2/\chi_0$, $X_{\rm r}/\chi_0$.

We expect the quantum fluctuations to provide tunnel coupling between the distinct minimums of the action. This will result in a non-degenerate ground state even at $X_{\rm r}=0$  that is a quantum superposition of two topologically distinct gapped states.

\section{Quantum corrections}
\label{sec:quantum}
In this Section, we consider the limit of $g \ll G/\ln G$ where we can average the nanostructure action and all quantities involved over the Gaussian fluctuations of $\chi$ produced by the external circuit (see Eq. \ref{eq:averagedaction}). Such averaging is impossible to do in general owing to the complexity of the resulting non-linear action. We restrict ourselves to the evaluation of the first-order correction.

A most straightforward way to proceed is to take the nanostructure action given by Eq. \ref{eq:actionrescaled1}, regard the time-dependent part of $X$, $x(\tau)$ as a perturbation entering the minimization equations for $\hat{U}$, solve those by subsequent iterations to the second order in $\delta{\chi}$,
\begin{align}
\hat{U} &= \hat{U}^{(0)} + \hat{U}^{(1)} + \hat{U}^{(1)};\\
\hat{U}^{(1)} &= \hat{A}\hat{x}\hat{B};\\
\hat{U}^{(2)} &= \hat{C}\hat{x}\hat{D} \hat{x}\hat{E},
\end{align} 
$\hat{U}^{(0)}, \hat{A} - \hat{E}$ being the operators diagonal in energy representation, average this over the fluctuations employing $\langle \langle x(\tau) x(\tau')\rangle \rangle = \lambda \delta(\tau-\tau')$. That results in correction 
\begin{equation}
\delta U(\tau -\tau') = \lambda \int d\tau_1  C(\tau-\tau_1) D(0) E(\tau_1 -\tau')
\end{equation}
to be compared with $U^{(0)}(\tau-\tau')$. 

We proceed in an equivalent, slightly more difficult but more instructive way. We take the nanostructure action given by Eq. \ref{eq:actionrescaled2} and substitute there the operators $\hat{U}, \hat{\tilde{\epsilon}}$ in a diagonal form plus non-diagonal deviations. We expand the action up to the second order in these deviations thereby accounting for the fluctuations of $X$ in this order, minimize the resulting quadratic action, and average over the fluctuations. This results in an additional $\propto \lambda$ term in the action, that is a functional in diagonal elements of the operators. Subsequent minimization over the diagonal elements permits to find the interaction correction to those.

To start, we rewrite the action employing the Lagrange multiplier $\hat{M}$ to ensure unitarity of $U$,
\begin{align}
{\cal S}_{\rm ns} &= -\frac{4 r^2 g}{2} \mathop{{\rm min}}_{\hat{U},\hat{M},\tilde{\epsilon}}[{\rm Tr} [ \hat{U}^\dagger \hat{A} +\hat{A}^\dagger\hat{U} \nonumber \\  &- (\hat{\epsilon}-\hat{\tilde{\epsilon}})^2 - \hat{M}(\hat{U} \hat{U}^\dagger -1)]
\end{align}
We separate the operators into diagonal and non-diagonal parts (we skip hats for diagonal operators) : 
$\hat{A} = A + \hat{a}$, $\hat{M} = M +\hat{M}$, $\hat{U} = U + \hat{u}$, $\hat{a} = \hat{x} + i \hat{p}$. The action up to the terms quadratic in diagonal elements reads:

\begin{align}
 {\cal S}_{\rm qd} &= -\frac{4 r^2 g}{2} {\rm Tr} [ \hat{u}^\dagger \hat{a} +\hat{a}^\dagger \hat{u} - \hat{p}^2 \nonumber\\
 &-\hat{m}( u U^\dagger + U \hat{u}^\dagger) - M \hat{u}\hat{u}^\dagger]
\end{align}
The minimization of this part with respect to all variables except $\hat{x}$ gives
\begin{align}
 {\cal S}_{\rm qd} &= -\frac{4 r^2 g}{2} \sum_{k,l} 
 \frac{|U_l - U_k^*|}{2 D_{kl}} |x_{kl}|^2,\\
 D_{kl} &\equiv (M_k -1)|U_k|^2 + (M_l -1)|U_l|^2 \\
 &+ \frac{1}{2}|U_l - U^*_k|^2 \nonumber
\end{align}
where $k,l$ index the discrete Matsubara energies. Here, $|U_k|^2 \ne 1$: although the matrix $\hat{U}$ is unitary, it also contains non-diagonal elements.
The averaging over the quantum fluctuations yields the quantum correction to the action, 
\begin{align}
\label{eq:actioncorrection}
{\cal S}_{\rm q} &= -\frac{4 r^2 g}{2} \lambda (k_BT)\sum_{k,l} \frac{|U_l - U_k^*|}{2 D_{kl}} \nonumber \\
&\equiv -\frac{4 r^2 g}{2} S_{\rm q} \\
\end{align}
To obtain the quantum corrections to the quantities, one has to minimize it with the diagonal part of the action,
\begin{align}
{\cal S}_{0} = -\frac{4 r^2 g}{2} \sum_{k} [U^*_k A_k + A^*_k U_k - (\epsilon_k - \tilde{\epsilon}_k)^2 \nonumber \\
+ M_k(|U_k|^2 -1).
\end{align}
The resulting minimization equations read ($U \equiv R+ i Y$ with real $R$, $Y$)
\begin{align}
X - MR + \frac{1}{2} \frac{\partial S_{\rm q} }{\partial R} &=0\label{eq:cor1}\\
1 - R^2 - Y^2 +\frac{1}{2} \frac{\partial S_{\rm q}}{\partial M}&=0 \label{eq:unitaritycorrection}\\
Y(1-M) +\epsilon +\frac{1}{2} \frac{\partial S_{\rm q}}{\partial Y}&=0 \label{eq:cor2}
\end{align}
with $\tilde{\epsilon} - \epsilon = Y$.

\begin{figure}[hbt!]
\begin{center}
	\includegraphics[width=0.8\columnwidth]{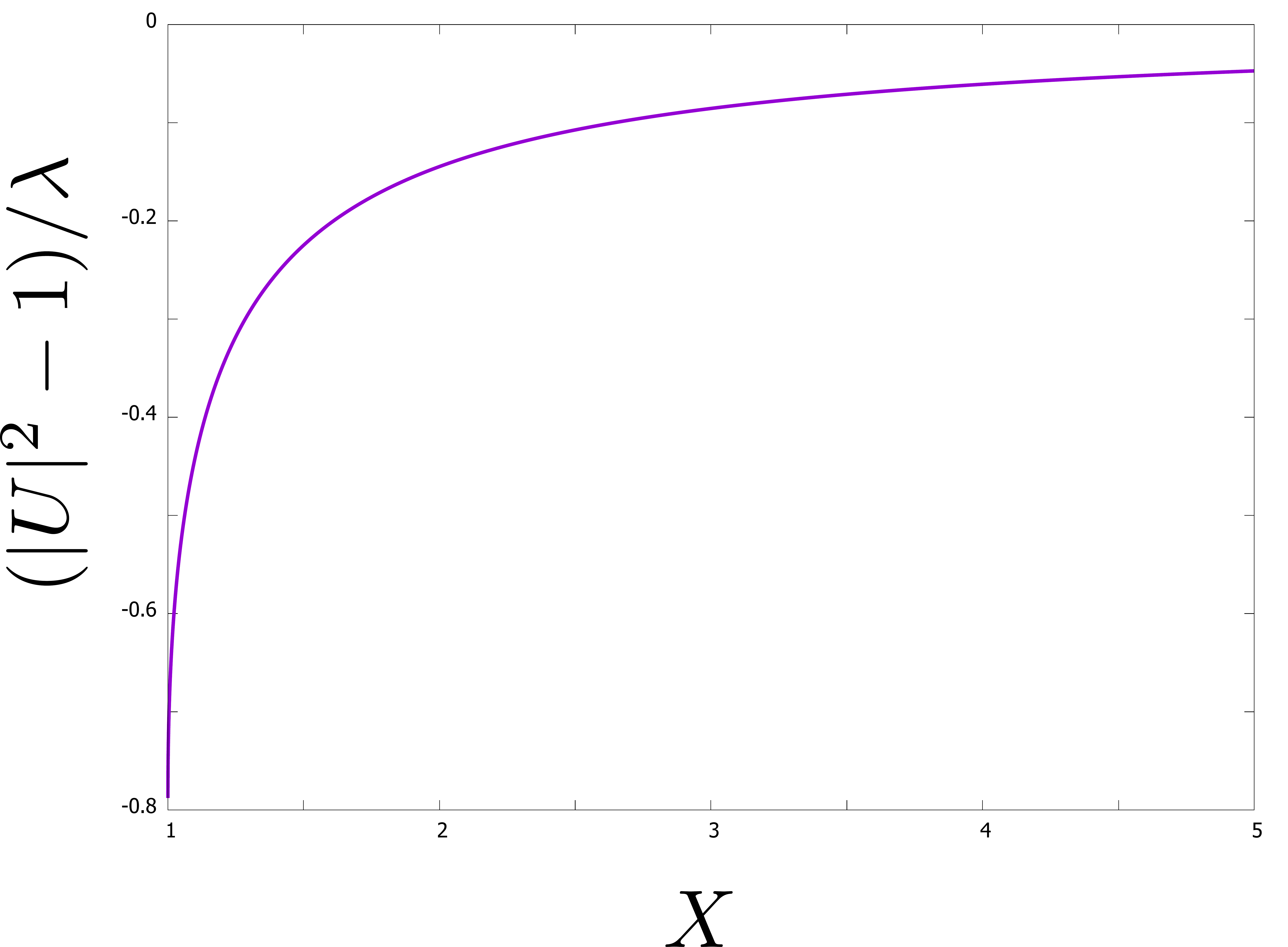} \\
	\caption{Interaction correction to "unitarity" of the Green function matrix versus $X$. The gapped phase, zero energy. The correction is of the order of $\lambda$ and remains finite at the critical point $X=1$. }\label{fig:correction}
	\end{center}	
\end{figure}

To give an example of practical calculation, let us evaluate a "correction to unitarity" given by Eq. \ref{eq:unitaritycorrection} that quantifies the importance of fluctuation-induced non-diagonal matrix elements in $U$. We compute $\partial{S_{\rm M}}$,
\begin{align}
\frac{\partial{S_{\rm q}}}{\partial M_l} = \frac{2(-Y_l Y_k +R_l R_k -1)}{(M_l +M_k -1 -R_kR_l +Y_l Y_l)^2}
\end{align}
We concentrate on zero energy and gapped phase, so we substitute $M_l=X, R_l=1, Y_l=0$ and $M_k =\sqrt{\tilde{\epsilon}^2 + X^2}$, $R_k = X/\sqrt{\tilde{\epsilon}^2 + X^2}$, $Y_k = \tilde{\epsilon}/\sqrt{\tilde{\epsilon}^2 + X^2}$ and change integration variable to $\tilde{\epsilon}$ as we did to derive Eq. \ref{eq:der}. This gives (see Fig. \ref{fig:correction} for the plot)
\begin{align}
&\frac{|U_0|^2 -1}{\lambda} =  \sqrt{\frac{X-1}{X+1}}\frac{2 X(X+3)}{(1+X)^2} \left( 1 -  \right. \nonumber\\
&  \left. \frac{{\rm arctan}(\sqrt{X^2-1})}{\pi} \right)+ \frac{8(2-X)}{3\pi(1+X)^2} -1 
\end{align} 
The correction remains finite at the critical point, $|U|^2 - 1 =\lambda (2/3\pi -1)$, although there is a square-root singularity in this point. At $X \to \infty (r \to 0)$ $|U|^2 - 1 =-\lambda(2/3\pi)/X$
In non-rescaled units, this implies that the correction amounts to 50 \% at $r=0$, $\chi = 4/3\pi G$, as shown in Fig \ref{fig:first} b.   

Let us compute the interaction-induced shift of the critical point that is located at $X=1$ for $\lambda=0$. We regard the total action as a function of four real parameters $y_a = (R, Y, M, \tilde{\epsilon})$ at zero energy. The critical point is determined from the condition ${\rm det} \partial_a \partial_b {\cal S} =0$ under constraint $\partial_a {\cal S}=0$. We find the matrix  $\partial_a \partial_b {\cal S}_0$, diagonalize it at the critical point $\lambda=0$  and bring in as small perturbations the shift of the parameters and the interaction correction action ${\cal S}_q$ . The condition of zero eigenvalue then reads
\begin{align}
\delta M = \frac{1}{2} \frac{\partial^2 {S}_{{\rm q}}}{\partial Y^2}
\end{align}
where $\delta M$ is contributed by $\delta X$ and the first-order correction computed from Eqs. \ref{eq:cor1}, \ref{eq:unitaritycorrection}, \ref{eq:cor2}. With this, the shift of the critical point
\begin{align}
\delta X = \frac{1}{2} \left(
\frac{\partial^2 S_{\rm q}}{\partial Y^2} -
\frac{\partial S_{\rm q}}{\partial R}
+\frac{\partial S_{\rm q}}{\partial M}
\right)
\end{align}
Substituting the values at the critical point to the derivatives of $S_{\rm q}$ and performing integration over $\tilde{\epsilon}$ yields
\begin{align}
\label{eq:shift}
\delta X = - \lambda \left(\frac{2}{3\pi} -1\right) \approx - 0.79 \lambda.
\end{align}
In non-rescaled coordinates, it corresponds to the shift of the phase boundary in the main axis direction by $\Delta \chi = \pm 0.79/ G$.

It is important to notice that the denominator in the expression for ${\cal S}_{\rm q}$, Eq. \ref{eq:actioncorrection}, vanishes at low energies in the gapless phase,
\begin{align}
D_{kl} \propto |\epsilon_k| + |\epsilon_l|,
\end{align}
provided the energies are of opposite sign, ${\rm sgn}(\epsilon_k) {\rm sgn}(\epsilon_l) = -1$. This form of the denominator manifests the existence of low-energy modes. The presence of low-energy modes might seem surprising: however, in the next Section we reveal that the gapless phase may be related to a breaking of a continuous symmetry, so this is just a manifestation of the Goldstone mechanism. This feature in the denominator might lead to low-energy divergences in the first-order interaction corrections under consideration. However, it does not: rather, the correction exhibit specific non-analytical terms in their low-energy expansion, 
\begin{equation}
A(\epsilon) = A(0) + \epsilon \ln(\epsilon_D/|\epsilon|)
\end{equation}
while this dependence is analytical in non-interacting quantities. 

To illustrate, we compute the correction to $Y$ given by
\begin{align}
Y^{(1)} =& \frac{2}{M-R^2} \left(
MY \frac{\partial S_{\rm q}}{\partial M}
-RY\frac{\partial S_{\rm q}}{\partial R}\right. \nonumber \\& \left.
+R^2\frac{\partial S_{\rm q}}{\partial Y}
\right)
\end{align}
At low energies of the opposite sign, the integrand takes the following form
\begin{align}
Y^{(1)}(\epsilon) &= \lambda \frac{X^2}{2(1-X^2)^{3/2}}\int_0^\infty \frac{d \epsilon'}{2\pi} \left[ 2 \frac{\epsilon -\epsilon'}{\epsilon +\epsilon'} \right. \nonumber\\&- \left.\frac{(\epsilon -\epsilon')^2}{(\epsilon +\epsilon')^2}\right]
\end{align}
This yields 
\begin{align}
Y^{(1)}(\epsilon) = {\rm const} + \frac{3 \lambda X^2}{4\pi(1-X^2)^{3/2}} \epsilon \ln\left[\frac{\epsilon_D}{|\epsilon|}\right],
\end{align}
$\epsilon_D \simeq 1$ being the cut-off energy.
We shall compare this with the low-energy dependence of $Y$ without interaction
\begin{equation}
Y(\epsilon) = \sqrt{1-X^2} + \epsilon \frac{X^2}{1-X^2}.
\end{equation}
This signals the break-down of the perturbation theory at arbitrary weak interaction: the energy dependence is dominated by correction at an exponentially small energy scale 
\begin{equation}
\label{eq:star}
\epsilon^{\star} \simeq \epsilon_D \exp\left(-\frac{\sqrt{1-X^2}}{ 3\lambda}\right).
\end{equation}
This energy scale increases upon approaching the critical point, see the discussion in the following Section.
\section{Near the boundary}
\label{sec:boundary}
In this Section, we derive a simplified action valid near the phase transition line separating the gapped and gapless phases. This action resembles the Landau Hamiltonian commonly incorporated for the description of the second-order phase transitions. However, the order parameter is a matrix with two time indices. We derive the proper scaling of the action.

It is convenient to start from the action in the form given by Eq. \ref{eq:actionrescaled1}. We note that at $X>0$ in the gapped phase and near the phase transition $\hat{\phi} \ll 1$ so we can expand $U$ in powers of $\hat{\phi}$. With this,
\begin{align}
\label{eq:Landau}
{\cal S}_{\rm ns}  = \frac{4 r^2 g}{2}\mathop{{\rm min}}_{\hat{\psi}} {\rm Tr}[ -2 \epsilon \psi + (a + \hat{\tilde{a}}) \frac{\psi^2}{2} + \frac{\psi^4}{4}] 
\end{align}
Here, $a \equiv (X-1)/2$ is the critical parameter of the second-order phase transition. If we neglect the fluctuations $\hat{\tilde{a}}$ and the term with $\epsilon$, the action describes the transition between the symmetric phase $\hat{\phi}=0$ at $a>0$ and the symmetry-broken phase $\hat{\psi}^2 = - a$. The symmetry-broken solution is highly degenerate if the eigenvalues of $\hat{\psi}$, $\pm \sqrt{-a}$, are of different sign: any unitary transformation  $\hat{psi} \to U^{-1} \hat{\psi} \hat{U}$ would produce a distinct solution of the same energy. The term with $\epsilon$ plays the role of a peculiar anisotropy term that breaks the degeracy and makes the solutions for $\psi$ unique on both sides of the transition.
Without fluctuations, $\psi$ is diagonal in energy, and $\psi(\epsilon) = -\psi(\epsilon)$.
Its equilibrium value is computed from $2\epsilon = a \psi +\psi^3$. At $a>0$ $\psi(\epsilon)$ is an analytical function of $\epsilon$ at $\epsilon \to 0$. This comforts the fact that it describes the gapped phase: the gap may be defined as the energy of the lowest singularity of $\psi(\epsilon)$ in the complex plane of $\epsilon$. Correspondinly, $\psi(\epsilon)$ is non-analytical in the gapless phase: $\psi(\epsilon) = {\rm sgn} (\epsilon) \sqrt{-a}$ at small energies.

We consider here the quantum limit where we can just average the action and $\psi$ over Gaussian fluctuations of $\tilde{a}$, $\langle \langle \tilde{a}(\tau) \tilde{a}(\tau') \rangle\rangle = (\lambda/4)\delta(\tau-\tau')$. The avegared $\langle\psi(\epsilon)\rangle$ become complex functions of energy, yet the transition point is defined in the same way, $\langle\psi(\epsilon)\rangle$ is an analytical function above the transition point and non-analytical otherwise.

Let us determine the scale of $a$ at which the fluctuations become important, the perturbation theory does not work, and deviations of $\langle\psi(\epsilon)\rangle$ form non-interacting limit are significant. Equating the scales of three terms in  Eq. \ref{eq:Landau}, we see that the scale $a_s$ determines the scale of $\psi$, $\psi_s = (a_s)^{1/2}$, and the energy, $\epsilon_s = a_s \psi_s = a_s^{3/2}$. The fluctuation of $a$ is estimated as $(\Delta a)^2 = \lambda \epsilon_s$. Equating this to $a$, we obtain $a_s = \lambda^2$. In non-rescaled coorditates, this reproduces the estimation $\delta \chi_s/\delta \chi_s = G^{-2}$ from the previous Section (see also Fig. \ref{fig:first} c.)

The scaling implies that 
\begin{equation}
\langle\psi(\epsilon)\rangle = a_s^{1/2} F(a/a_s, \epsilon/a_s^{3/2}) 
\end{equation}
The non-interacting values are reproduced at large values of the arguments of this scaling function. For practical applications, one also needs to account for the shift of the transition point (Eq. \ref{eq:shift}) that comes from the higher-energy fluctuations and does not conform this scaling which takes place in a small vicinity of the $shifted$ transition point.

Let us note that the exponentially small low-energy scale $\epsilon^\star$ found in the previous Section (Eq. \ref{eq:star}) also conforms to this scaling. Near the critical point, but at $|a| \gg a_s$ it can be expessed as
\begin{align}
\epsilon^{\star} \simeq a_s^{3/2} (a/a_s)^{3/2} \exp\left(-\frac{2}{3}\left(\frac{|a|}{a_s}\right)^{1/2} \right).
\end{align}

So it becomes of the order of all other scales at $a \simeq a_s$.
This provokes a hypothesis that we formulate in the concluding Section.
 
\section{Conclusions and hypotheses}
\label{sec:conclude}

There are few examples of condensed matter models where an arbitrary weak interaction qualitatively changes the fermionic spectrum, the superconductivity is the most famous one. In this paper, we predict a drastic effect of weak interaction on the Andreev spectrum near the special points in multi-terminal semiclassical superconducting nanostructures. 

This is a generic effect to arise in any nanostructure. Our approach is valid for normal nanostructures but also for superconducting ones, provided they are smaller than the superconducting coherence length. It can be experimentally observed by studying the tunnelling to these nanostructures at low energies.

We have developed a general interaction model and came up with a simple universal action describing the situation. This is a complex non-local and non-linear quantum field theory that cannot be analytically treated by existing methods. Its numerical study is plausible but requires a significant effort in view of the matrix nature of the order parameter. In this paper, we analytically studied a semiclassical limit and the first-order interaction correction in the quantum limit. 

In both limits, our results indicate that the effect on the spectrum is drastic in the close vicinity of the special point at the scale defined by interaction. The domain of the gapless phase is squeezed. In semiclassical limit, we see the failure of the topological protection: two gapped phases of distinct topology come to contact not being separated by a gapless phase.

We would like to put forward two hypotheses to be confirmed or disproved in the course of  further research. First hypothesis is that the gapped phases are in direct contact at the special point also in the quantum limit. This hypothesis is based on continuity: one can go from semiclassical to quantum limit by changing the parameter $g$. An alternative would be a phase transition upon this change.

The second hypothesis is provoked by an exponentially small low-energy scale $\epsilon^\star$ found in the gapless phase. It may be that a gap develops at this energy scale, and the gapless phase without interaction always becomes gapped in the presence of an arbitrary weak interaction, even far from the special point. In this case, the phase transition between gapped and gapless phase would become a crossover not separating distinct phases. The fact that $\epsilon^\star$ conforms to the critical scaling supports this hypothesis.

To support open science and open software initiatives and to comply with institutional policies, we have published all relevant code and instructions for running it on the Zenodo repository \cite{DrasticCode}. 
 
\acknowledgments

This research was supported by the European
Research Council (ERC) under the European Union's
Horizon 2020 research and innovation programme (grant
agreement No. 694272).

\bibliography{bibliography}

\end{document}